# Feature Engineering-Based Detection of Buffer Overflow Vulnerability in Source Code Using Neural Networks

Mst Shapna Akter[1], Hossain Shahriar[2], Juan Rodriguez Cardenas[3], Sheikh Iqbal Ahamed[4], and Alfredo Cuzzocrea[5]

[1]Department of Computer Science, Kennesaw State University, USA
[2]Department of Information Technology, Kennesaw State University, USA
[3]Department of Information Technology, Kennesaw State University, USA
[4]Department of Computer Science, Marquette University, USA
[5]iDEA Lab, University of Calabria, Rende, Italy

*Abstract*—One of the most significant challenges in the field of software code auditing is the presence of vulnerabilities in software source code. Every year, more and more software flaws are discovered, either internally in proprietary code or publicly disclosed. These flaws are highly likely to be exploited and can lead to system compromise, data leakage, or denial of service. To create a large-scale machine learning system for function-level vulnerability identification, we utilized a sizable dataset of C and C++ open-source code containing millions of functions with potential buffer overflow exploits. We have developed an efficient and scalable vulnerability detection method based on neural network models that learn features extracted from the source codes. The source code is first converted into an intermediate representation to remove unnecessary components and shorten dependencies. We maintain the semantic and syntactic information using state-of-the-art word embedding algorithms such as GloVe and fastText. The embedded vectors are subsequently fed into neural networks such as LSTM, BiLSTM, LSTM-Autoencoder, word2vec, BERT, and GPT-2 to classify the possible vulnerabilities. Furthermore, we have proposed a neural network model that can overcome issues associated with traditional neural networks. We have used evaluation metrics such as F1 score, precision, recall, accuracy, and total execution time to measure the performance. We have conducted a comparative analysis between results derived from features containing a minimal text representation and semantic and syntactic information. We have found that all neural network models provide higher accuracy when we use semantic and syntactic information as features. However, this approach requires more execution time due to the added complexity of the word embedding algorithm. Moreover, our proposed model provides higher accuracy than LSTM, BiLSTM, LSTM-Autoencoder, word2vec and BERT models, and the same accuracy as the GPT-2 model with greater efficiency.

## I. INTRODUCTION

Security in the digital realm is becoming increasingly important, but there is a significant threat to cyberspace from invasion. Attackers can breach systems and applications due to security vulnerabilities caused by hidden software defects. Internally, proprietary programming contains thousands of these flaws each year [1]. For example, the ransomware Wannacry swept the globe by using a flaw in the Windows server message block protocol [2]. According to the Microsoft Security Response Center, there was an industry-wide surge in high-severity vulnerabilities of 41.7% in the first half of 2015. This represents the greatest proportion of software vulnerabilities in at least three years, accounting for 41.8% [3]. Furthermore, according to a Frost and Sullivan analysis released in 2018, severe and high severity vulnerabilities increased from 693 in 2016 to 929 in 2017, with Google Project Zero coming in second place in terms of disclosing such flaws. On August 14, 2019, Intel issued a warning about a high-severity vulnerability in the software it uses to identify the specifications of Intel processors in Windows PCs [4]. The paper claims that these defects, including information leaking and denial of service assaults, might substantially affect software systems. Although the company issued an update to remedy the problems, attackers can still use these vulnerabilities to escalate their privileges on a machine that has already been compromised. In June 2021, a vulnerability in the Windows Print Spooler service was discovered that allowed attackers to execute code remotely. The vulnerability, known as PrintNightmare, was caused by a buffer overflow and affected multiple versions of Windows in 2021 [5]. Microsoft released a patch to address the issue, but reports later emerged that the patch was incomplete and still left systems vulnerable.

To reduce losses, early vulnerability detection is a good technique. The proliferation of open-source software and code reuse makes these vulnerabilities susceptible to rapid propagation. Source code analysis tools are already available; however, they often only identify a small subset of potential problems based on pre-established rules. Software vulnerabilities can be found using a technique called vulnerability detection. Conventional vulnerability detection employs static and dynamic techniques [6]. Static approaches evaluate source code or executable code without launching any programs, such as data flow analysis, symbol execution [7], and theorem proving [8]. Static approaches can be used early in software development and have excellent coverage rates, but they have a significant false positive rate. By executing the program, dynamic approaches like fuzzy testing and dynamic symbol execution can confirm or ascertain the nature of the software. Dynamic methods depend on the coverage of test cases, which results in a low recall despite their low false positive rate and ease of implementation. The advancement of machine learning technology incorporates new approaches to address the limitations of conventional approaches. One of the key research directions is to develop intelligent source code-based vulnerability detection systems. It can be divided into three categories: using software engineering metrics, anomaly detection, and weak pattern learning [9]. Initially, software engineering measures, including software complexity [10], developer activity [11], and code commits [12] were investigated to train a machine learning model. This strategy was motivated by the idea that software becomes more susceptible as it becomes more complicated, but accuracy and recall need to be improved. Allamanis et al. [13] have shown that the syntactic and semantic information in the code increases the detection accuracy in anomaly detection. Moreover, one work has shown the detection of the anomaly using fully-fledged codes [14]. It reveals previously unidentified weaknesses, but false positive and false negative rates are high. Another work has shown an approach with clean and vulnerable samples to learn vulnerable patterns [15]. This method performs very well but relies on the quality of the dataset. In our work, we propose a solution for detecting software buffer overflow vulnerability using neural networks such as Simple RNN, LSTM, BiLSTM, word2vec, BERT, GPT2, and LSTM-Autoencoder. We first transform source code samples into the minimum intermediate representations through a tokenizer provided by the Keras library. Later, we extract semantic features using word embedding algorithms such as GloVe and fastText. After finishing the data preprocessing stage, we feed the input representation to the neural networks for classification. Moreover, we develop a neural network that works best among all the models. All the models have been evaluated using evaluation metrics such as f1 score, precision, recall, accuracy, and total execution time. The following is a summary of our contributions:

1. Extracting semantic and syntactic features using GloVe and fastText. 2. Vulnerability Detection in Source Code using LSTM, BiLSTM, LSTM-Autoencoder, word2vec, BERT, and GPT-2 with an minimal intermediate feature representation of the texts. 3. Vulnerability Detection in Source Code using LSTM, BiLSTM, LSTM-Autoencoder, word2vec, BERT, and GPT-2 with semantic and syntactic features. 4. Proposal of a neural network that outperforms the results derived from existing models. Comparison between results derived from neural networks trained with a minimal intermediate feature representation of the texts and semantic and syntactic features.

The rest of the paper is organized as follows: we provide a brief background study on software vulnerability detection in section 2. Then we explain the methods we followed for our experimental research in section 3. The results derived from the experiment are demonstrated in Section 4. Finally, section 5 concludes the paper.

II. LITERATURE REVIEW

Researchers are interested in the recently developed machine learning strategy for identifying and preventing software and cybersecurity vulnerabilities in order to address the shortcomings of conventional static and dynamic code analysis techniques. Various machine learning techniques, including naive bayes, logistic regression, recurrent neural networks (RNN), decision trees, and support vector machines are successfully used for classifying software security activities like malware, ransomware, and network intrusion detection. We have examined machine learning-related papers that have been applied to the software security domain. Previously, Zeng et al. [16] reviewed software vulnerability analysis and discovery using deep learning techniques. They found four game-changing methods that contributed most to software vulnerability detection using deep learning techniques. These concepts are automatic semantic feature extraction using deep learning models, end-to-end solutions for detecting buffer overflow vulnerabilities, applying a bidirectional Long Short-Term Memory (BiLSTM) model for vulnerability detection, and deep learning-based vulnerability detectors for binary code. Zhou et al. [17] proposed a method called graph neural network for vulnerability identification with function-level granularity to address the issue of information loss during the representation learning process. They transformed the samples into a code property graph format. Then, a graph neural network made up of a convolutional layer and a gated graph recurrent layer learned the vulnerable programming pattern. This method improves the detection of intra-procedural vulnerabilities. However, they did not address inter-procedural vulnerabilities. Iorga et al. [18] demonstrated a process for early detection of cyber vulnerabilities from Twitter, building a corpus of 650 annotated tweets related to cybersecurity articles. They used the BERT model and transfer learning model for identifying cyber vulnerabilities from the articles. The BERT model shows 91% accuracy, which they found adequate for identifying relevant posts or news articles. Sauerwein et al. [19] presented an approach for automated classification of attackers' TTPs by combining NLP with ML techniques. They extracted the attackers' TTPs from unstructured text. To extract the TTPs, they used a combination of NLP with ML techniques. They assessed all potential combinations of the specified NLP and ML approaches with 156 processing pipelines and an automatically generated training set. They found that tokenization, POS tagging, IoC replacement, lemmatization, one-hot encoding, binary relevance, and support vector machine performed best for the classification of techniques and tactics. Harer et al. [20] created a dataset composed of millions of open-source functions annotated with results from static analysis. The performance of source-based models is then compared against approaches applied to artifacts extracted from the build process, with source-based methods coming out on top. The best performance is found when combining characteristics learned by deep models with tree-based models. They evaluated the use of deep neural network models alongside more conventional models like random forests. Finally, their best model achieved an area under the ROC curve of 0.87 and an area under the precision-recall curve of 0.49. Pistoia et al. [21] surveyed static analysis methods for identifying security vulnerabilities in software systems. They discussed three topics that have been linked to security vulnerability sources: application programming interface conformance, information flow, and access control. They addressed static analysis methods for stack-based access control and role-based access control separately since access control systems can be divided into these two main types. They reviewed some effective static analysis techniques, including the Mandatory Access Rights Certification of Objects (MARCO) algorithm, the Enterprise Security Policy Evaluation (ESPE) algorithm, the Static Analysis for Validation of Enterprise Security (SAVES) algorithm, and Hammer, Krinke, and Snelting's algorithm. However, static analysis produces false positive results and relies on predefined rules. For new errors, the static analysis method is unsuitable, as it cannot recognize and detect them.

## III. METHODOLOGY

From the standpoint of source code, the majority of flaws originate in critical processes that pose security risks, such as functions, assignments, or control statements. Adversaries can directly or indirectly affect these crucial operations by manipulating factors or circumstances. To successfully understand patterns of security vulnerabilities from code, neural network models must be trained on a large number of instances. In this study, we analyze the lowest level of codes in software package functions, capable of capturing vulnerable flows. We utilized a sizable dataset containing millions of function-level examples of C and C++ code from the SATE IV Juliet Test Suite, the Debian Linux distribution, and open-source Git repositories on GitHub, as mentioned in Russell's work [22]. Our project employs the CWE-119 vulnerability feature, which indicates issues related to buffer overflow vulnerability. Buffer overflow occurs when data written to a buffer exceeds its length, overwriting storage units outside the buffer. According to a 2019 Common Weakness Enumeration report, buffer overflow vulnerability has become the most adversely affected issue. Although we focus on buffer overflow, our method can identify other vulnerabilities. Figure 1 illustrates an intra-procedural buffer overflow vulnerability. Our dataset is divided into three subfolders—train, validation, and test—each containing a CSV file with 100,000, 20,000, and 10,000 data instances, respectively. The CSV files store text data and corresponding labels, allowing systematic evaluation of the model's performance and adaptability throughout the learning process.

```
1  void bar(char *buf, char *src) {
2      strcpy(buf, src);
3  }
4  int main() {
5      char buf[10];
6      char src[10];
7      memset(src, 'A', 10);
8      src[10 - 1] = '\0';
9      bar(buf, src);
10     for (int i = 0; i <= 10; i++)
11     //writes buf [10] and overruns memory
12         buf[i] = 'B';
13     return 0;
14 }
```

Fig. 1: An example of buffer overflow vulnerability.

We analyzed the dataset and found some common words (shown in Table 1) with their corresponding counts. The visualization of common words in the dataset provides a preliminary understanding of what kind of important features the dataset might have.

TABLE I: Most common words and their frequencies

| index | Common_words | Count |
|---|---|---|
| 0 | = | 505570 |
| 1 | if | 151663 |
| 2 | {\n | 113301 |
| 3 | == | 92654 |
| 4 | return | 77438 |
| 5 | * | 71897 |
| 6 | the | 71595 |
| 7 | }\n | 63182 |
| 9 | int | 53673 |
| 10 | /* | 51910 |
| 11 | i | 43703 |
| 12 | */\n | 43591 |
| 13 | + | 41855 |
| 14 | to | 39072 |
| 15 | && | 36180 |
| 16 | for | 35849 |
| 17 | }\n\n | 34017 |
| 18 | char | 33334 |
| 19 | else | 31358 |

*1) Data Preprocessing:* In this study, we conducted a series of data preprocessing techniques to prepare our dataset for the neural networks. The data preprocessing steps we employed include tokenization, stop word removal, stemming, lemmatization, and the use of pre-trained embeddings. Initially, we performed tokenization, which is the process of breaking down the source code into smaller units called tokens. Tokens represent the basic units of analysis for computational purposes in natural language processing tasks. For this process, we utilized the Keras tokenizer, which provides methods such as tokenize() and detokenize() to process plain text and separate words [23]. Following tokenization, we applied stop word removal, stemming, and lemmatization techniques to further preprocess the tokens. Stop word removal eliminates common words that do not provide significant information, while stemming and lemmatization normalize the tokens by reducing them to their root form. These techniques help in reducing noise and improving the efficiency of the neural networks.

We first converted the tokens into numerical representations using minimal intermediate representation with the Keras tokenizer. The Keras tokenizer assigns a unique integer index to each token in the vocabulary and represents the source code as a sequence of these integer indices. This representation is more efficient than one-hot encoding, as it does not involve creating large, sparse vectors. However, it still lacks semantic information about the tokens. To further enhance the representation of the source code tokens and better capture semantic and syntactic information, we utilized pre-trained embeddings, namely GloVe and fastText. We stacked GloVe and fastText embeddings together for extracting the semantic and syntactic information from the source code. Both of these embeddings have demonstrated strong performance in various NLP tasks and can effectively capture the relationships between words in the source code. GloVe is an unsupervised learning algorithm that generates vector representations of words based on global word-word co-occurrence statistics from a corpus [24]. FastText, an extension of the skip-gram method, generates character n-grams of varying lengths for each word and learns weights for each n-gram, as well as the entire word token, allowing the model to capture the meaning of suffixes, prefixes, and short words [25]. We separately fed the minimal intermediate representation with Keras tokenizer and the semantic and syntactic representations derived from GloVe and fastText into our neural network models. This approach allowed us to compare the performance of the models when using different input representations, helping us identify the most effective method for detecting security vulnerabilities in the source code.

*A. Classification Models*

In this section, we discuss various classification models that were utilized in our study. These models include Simple RNN, LSTM, BiLSTM, LSTM-Autoencoder, Word2vec, BERT, and GPT-2. These models are designed to work with different types of data, such as text, time series, and sequences, and have been widely employed in natural language processing and other related tasks.

*B. Simple Recurrent Neural Network (RNN)*

The Simple Recurrent Neural Network (RNN) is a type of artificial neural network that can model sequential data by utilizing a directed graph and temporally dynamic behavior. RNNs consist of an input layer, a hidden layer, and an output layer [26]. These networks have a memory state added to each neuron, allowing them to capture temporal dependencies in the data. The dimensionality of the input layer in our Simple Recurrent Neural Network (RNN) model is determined based on the input data features. The hidden layer consists of 256 units, which use memory states to capture temporal dependencies in the data. We use the hyperbolic tangent (tanh) activation function in the hidden layer to introduce non-linearity into the model. We chose this activation function due to its ability to handle vanishing gradients more effectively compared to other activation functions like sigmoid. The output layer of the Simple RNN model is designed to generate predictions based on the processed input data. The number of units in the output layer corresponds to the number of classes, which is two. We use an appropriate activation function, such as sigmoid for binary classification, in the output layer to generate probability scores for each class. To optimize the model, we choose the Binary Cross entropy loss function and employ the Adam optimization algorithm. We set hyperparameters such as learning rate to 0.001, batch size to 32, and the number of training epochs to 50.

*C. Long short-term memory (LSTM)*

The Long Short-Term Memory (LSTM) is a type of recurrent neural network designed to solve the vanishing and exploding gradient problem of traditional RNNs.It was first proposed by Hochreiter and Schmidhuber [27]. Using this model for sequential datasets is effective, as it can handle single data points. It follows the Simple RNN model's design and is an extended version of that model [28, 29]. Our LSTM model consists of an input layer that determines the dimensionality of the input data features. We incorporated three hidden layers, each containing 128 memory cells that can capture long-term dependencies in the input sequence. The output of each LSTM layer is fed into a dropout layer with a dropout rate of 0.2 to prevent overfitting. The final output of the last LSTM layer is fed into a dense layer with two units and a sigmoid activation function to produce the final binary classification output. The LSTM cell comprises three gates: the input gate, forget gate, and output gate, which regulate the flow of information into and out of the cell. To introduce non-linearity into the model, we use the hyperbolic tangent (tanh) activation function in the LSTM cell. Furthermore, we utilize the Rectified Linear Unit (ReLU) activation function in the output layer to generate non-negative predictions. We optimize the LSTM model using the Binary Cross-Entropy loss function and Adam optimization algorithm. The model's hyperparameters include a learning rate of 0.001, batch size of 32, and 50 training epochs.

*D. Bidirectional Long short-term memory (BiLSTM)*

The Bidirectional Long Short-Term Memory (BiLSTM) is a type of recurrent neural network that enhances the capabilities of the traditional LSTM by introducing bidirectional processing of the input sequence. It was first proposed by Graves [30]. This idea sets it apart from the LSTM model, which can learn patterns from the past to the future [31] .Our BiLSTM model comprises an input layer that determines the dimensionality of the input data features. We have incorporated three hidden layers, each containing 128 memory cells that can capture long-term dependencies in the input sequence. The output of each BiLSTM layer is fed into a dropout layer with a dropout rate of 0.2 to prevent overfitting. The final output of the last BiLSTM layer is fed into a dense layer with two units and a sigmoid activation function to produce the final binary classification output. The BiLSTM cell has two sets of three gates, namely the input gate, forget gate, and output gate, one set that processes the input sequence in the forward direction and another set that processes the input sequence in the backward direction. This bidirectional processing allows the model to capture dependencies in both the past and future context of the input sequence. To introduce non-linearity into the model, we use the hyperbolic tangent (tanh) activation function in the BiLSTM cell. Furthermore, we utilize the Rectified Linear Unit (ReLU) activation function in the output layer to generate non-negative predictions. We optimize the BiLSTM model using the Binary Cross-Entropy loss function and Adam optimization algorithm. The model's hyperparameters include a learning rate of 0.001, batch size of 32, and 50 training epochs.

*E. LSTM-Autoencoder*

The LSTM-Autoencoder is a variant of the Long Short-Term Memory (LSTM) model that utilizes an autoencoder

architecture. The LSTM-Autoencoder is designed to read input sequences, encode sequences, decode sequences, and reconstruct sequences for a given sequential dataset, referred to as encoder-decoder [32]. Its performance is estimated based on how well the model can recreate the sequence. LSTM autoencoder can be used on video, text, audio, and time-series sequence data. The model accepts a series of various lengths of inputs and outputs for various purposes, such as translating from one language to another. The series is transformed into a vector representation by the encoder, and the vector is transformed back into a sequence of outputs or texts by the decoder. The meaning of the outputs is maintained in the vector representation. In this model, we have an input layer that determines the dimensionality of the input data features. The LSTM encoder layer contains 128 memory cells that can capture long-term dependencies in the input sequence. The LSTM decoder layer has the same number of memory cells as the encoder layer, which allows the model to reconstruct the input sequence. To introduce non-linearity into the model, we use the hyperbolic tangent (tanh) activation function in the LSTM cells. Additionally, we utilize the Mean Squared Error (MSE) loss function to calculate the reconstruction loss of the autoencoder. The model's hyperparameters include a learning rate of 0.001, batch size of 32, and 50 training epochs. To evaluate the performance of the LSTM-Autoencoder, we calculate the reconstruction error between the input and reconstructed sequence. The lower the reconstruction error, the better the model's ability to capture the input sequence's structure.

*F. Word2vec*

Word2vec is a word embedding model specifically designed for working with textual data. Word embedding is a technique for representing words that allows computer programs to understand words with similar meanings. By employing a neural network model to map words into vectors of real numbers, word2vec is capable of capturing significant accurate syntactic and semantic word relationships. After training, the two-layer neural network can recognize synonymous terms and suggest new words for incomplete phrases [33]. Our Word2vec model comprises an input layer that takes in the one-hot encoded words and a single hidden layer containing a specified number of neurons, which represent the latent dimensions of the word embeddings. We utilize the Skip-gram architecture with negative sampling to train the Word2vec model. In this architecture, the model predicts the surrounding words given a target word or predicts the target word given surrounding words. The negative sampling technique helps to efficiently train the model by reducing the computation required to update the weights of the model. The output layer is not used in the Word2vec model, and the trained weights of the hidden layer represent the learned word embeddings. These embeddings can be used in various downstream NLP tasks such as text classification, sentiment analysis, and machine translation. To optimize the model, we use the Stochastic Gradient Descent (SGD) optimization algorithm with an initial learning rate of 0.025 and decrease the learning rate linearly over time to 0.001.

We set the batch size to 128 and the number of training epochs to 5.

*G. BERT*

BERT (Bidirectional Encoder Representations from Transformers) is a state-of-the-art pre-trained language model developed by Google. BERT is a bidirectional transformer-based architecture that can capture the context of a word in a sentence by looking at the surrounding words [34]. The BERT model consists of 12 transformer blocks for the base version and 24 transformer blocks for the large version. Each transformer block has a multi-head attention mechanism and a feed-forward neural network, making it capable of modeling long-term dependencies in the input sequence. In our implementation of BERT, we utilized the pre-trained BERT model and fine-tuned it on our specific NLP task. We utilized the pre-trained BERT model with 12 transformer blocks, 12 attention heads, and 110 million parameters. We added a dense layer with 2 units and a sigmoid activation function to perform binary classification. We utilized the Binary Cross-Entropy loss function and Adam optimization algorithm to optimize the model. We set the learning rate to 2e-5 and the batch size to 32. To fine-tune the pre-trained BERT model, we trained it on our specific NLP task using a training set of 100,000 instances and a validation set of 20,000 instances. We trained the model for 3 epochs and evaluated its performance on a separate test set, which constists of 10,000 instances.

*H. GPT-2*

GPT-2 (Generative Pre-trained Transformer 2) is a state-of-the-art language model developed by OpenAI. It is a transformer-based language model that can generate coherent and fluent text in a wide range of styles and topics [35]. GPT-2 has a large number of parameters, with the base version having 117 million parameters, and the largest version having 1.5 billion parameters. In our implementation of GPT-2, we utilized the pre-trained GPT-2 model to generate text for our specific NLP task. We fine-tuned the pre-trained GPT-2 model on a large corpus of text relevant to our task to improve its performance. We used the GPT-2 model with 117 million parameters for our task. To fine-tune the pre-trained GPT-2 model, we used a training set of 100,000 instances and a validation set of 20,000 instances. We fine-tuned the model for 3 epochs and evaluated its performance on a separate test set. We used the perplexity metric to evaluate the performance of the model. We utilized the Adam optimization algorithm with a learning rate of 1e-5 and a batch size of 32 to optimize the model.

*I. Proposed Model*

We propose a stacking ensemble learning approach to improve the performance of our system. Stacking ensemble is an advanced machine learning technique that combines multiple heterogeneous weak learners (base models) to form a single stronger model (meta-learner). In this approach, the base models' predictions are used as input to the meta-learner, which ultimately makes the final prediction. The meta-learner used in this case is a logistic regression model, while the base models consist of Simple RNN, LSTM, BiLSTM, word2vec, and LSTM-Autoencoder. These models are trained with one-dimensional data as input.

Since the predicted dataset from Level 0 already contains the expected values' probabilities, the meta-learner can provide accurate probabilities from Level 0. To mitigate overfitting, the meta-learner is trained using both the validation dataset and the outputs. The final result is the level-1 prediction.

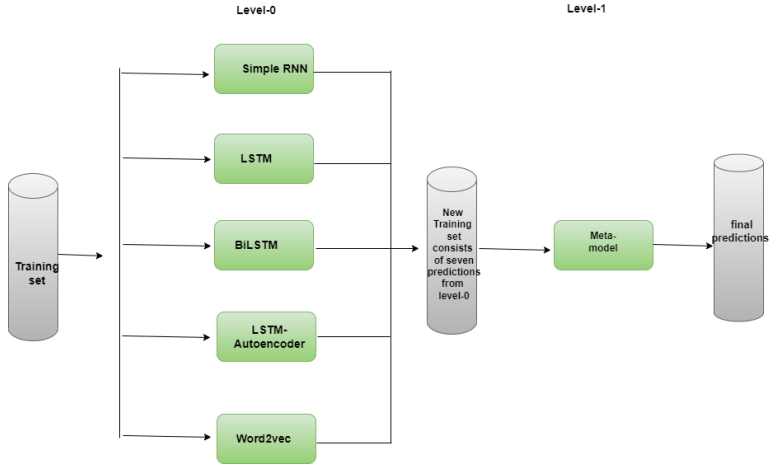

Fig. 2: Proposed stacking ensemble learning architecture.

The architecture is divided into two levels, Level 0 and Level 1, as shown in Figure 2. Level 0 consists of Simple RNN, LSTM, BiLSTM, word2vec, and LSTM-Autoencoder models. After learning the data patterns, each of the base models generates predictions simultaneously. All models in Level 0 contribute equally to the overall model performance.

Level 1, also referred to as the meta-learner, is built using logistic regression. The meta-learner at Level 1 is fed the Level 0 predicted outputs as input. Based on the Level 0 predictions, the meta-learner calculates the best weighted outputs. A "meta-learner" is a model that can quickly learn a pattern or adapt to different datasets with a small amount of training data. It learns patterns from the outputs generated by the five base models. As a result, the model can effectively learn completely new data and produce acceptable output. The meta-learner's parameters are a combination of the parameters of the five neural networks in the base models.

Mathematically, the stacking ensemble learning approach can be represented as follows:

Let $M$ be the number of base models, $p_i^m$ be the probability of the positive class for the $i$-th sample predicted by the $m$-th base model, and $w_m$ be the weight assigned to the $m$-th base model. The weighted probability $p_i^{weighted}$ for the $i$-th sample can be computed as:

$$p_i^{weighted} = \sum_{m=1}^{M} w_m \cdot p_i^m$$

The weights $w_m$ are determined by the meta-learner using the Level 0 predictions and the validation data. The final prediction $y_i^{final}$ for the $i$-th sample can be computed using the logistic function:

$$y_i^{final} = \frac{1}{1 + e^{-(p_i^{weighted})}}$$

By using a diverse set of base models, we can mitigate the limitations of traditional stacking ensemble approaches that employ similar base models, leading to similar predictions. If a single base model performs poorly on the dataset, there is a high likelihood that the final output will also be inferior. Conversely, with a diverse set of base models, the strengths and weaknesses of individual models complement each other, which results in a more robust and accurate overall model. This is because each base model is able to capture different aspects or patterns in the data, thereby reducing the reliance on any single model's performance. Additionally, the meta-learner can effectively combine these diverse predictions to generate a more accurate and stable final prediction, minimizing the impact of individual model biases or errors. In conclusion, the utilization of heterogeneous base models in a stacking ensemble approach provides a more resilient and powerful predictive model, capable of handling various types of data and delivering superior performance compared to traditional ensemble methods.

**Algorithm 1** Proposed Stacking Ensemble Learning Algorithm.

**Function** stacking_ensemble(*data, train_ratio, val_ratio, test_ratio*)

```
// Initialize Level 0 base models
simple_rnn ← SimpleRNN();
    lstm ← LSTM();
    bi_lstm ← BiLSTM();
    lstm_autoencoder ← LSTM_Autoencoder();
    word2vec_model ← Word2Vec();
    models ← [simple_rnn, lstm, bi_lstm,
    lstm_autoencoder, word2vec_model];
    // Initialize Level 1 meta-learner
meta_learner ← LogisticRegression()

// Split the data into training,
    validation, and testing sets
X_train, X_val, X_test, y_train, y_val, y_test ←
 data_split(data, train_ratio, val_ratio, test_ratio)

// Train Level 0 base models
foreach model in models do
 │ model.fit(X_train, y_train)
// Make predictions with Level 0 base
    models
Level0_outputs ← list()
  foreach model in models do
   │ pred ← model.predict(X_val)
   │  Level0_outputs.append(pred)
// Concatenate Level 0 outputs
Level0_outputs_combined ← concate-
 nate(Level0_outputs)

// Train Level 1 meta-learner
meta_learner.fit(Level0_outputs_combined, y_val)

// Make final predictions with Level 1
    meta-learner
Level0_test_outputs ← list()
  foreach model in models do
   │ test_pred ← model.predict(X_test)
   │  Level0_test_outputs.append(test_pred)
// Concatenate Level 0 test outputs
Level0_test_outputs_combined ← concate-
nate(Level0_test_outputs)

// Generate Level 1 final predictions
final_predictions ← meta_learner.predict(Level0_test_outputs)
return final_predictions
```

## IV. EVALUATION METRICS

In order to assess the performance of the Neural Networks and our proposed stacking ensemble model, we have employed a range of evaluation metrics that provide insight into various aspects of model performance. These metrics include precision, recall, F1-score, accuracy, and execution time. Each of these metrics contributes to a comprehensive understanding of the model's effectiveness, generalization, and efficiency [36–38]. Below, we provide a brief description of each evaluation metric:

### A. Precision

Precision is a measure of the accuracy of the positive predictions made by the model. It is calculated as the ratio of true positive predictions to the sum of true positive and false positive predictions. In other words, it quantifies the proportion of correct positive predictions among all the instances predicted as positive. A higher precision value indicates that the model is better at identifying relevant instances and minimizing false positive predictions.

$$\text{Precision} = \frac{\text{True Positives}}{\text{True Positives} + \text{False Positives}} \quad (1)$$

### B. Recall

Recall, also known as sensitivity or true positive rate, measures the proportion of actual positive instances that are correctly identified by the model. It is calculated as the ratio of true positive predictions to the sum of true positive and false negative predictions. A higher recall value indicates that the model is better at detecting positive instances and minimizing false negative predictions.

$$\text{Recall} = \frac{\text{True Positives}}{\text{True Positives} + \text{False Negatives}} \quad (2)$$

### C. F1-score

F1-score is the harmonic mean of precision and recall, and it provides a balanced measure of both metrics. It is particularly useful when dealing with imbalanced datasets, where one class is significantly more prevalent than the other. The F1-score ranges from 0 to 1, with a higher value indicating better overall performance of the model in terms of both precision and recall.

$$\text{F1-score} = 2 \cdot \frac{\text{Precision} \cdot \text{Recall}}{\text{Precision} + \text{Recall}} \quad (3)$$

### D. Accuracy

Accuracy is a widely-used metric that quantifies the proportion of correct predictions made by the model, both positive and negative, relative to the total number of instances. It provides an overall indication of the model's performance, but it may not be a reliable metric when dealing with imbalanced datasets, as it can be biased towards the majority class.

$$\text{Accuracy} = \frac{\text{True Positives} + \text{True Negatives}}{\text{Total Instances}} \quad (4)$$

### E. Execution Time

Execution time is a measure of the computational efficiency of the model. It refers to the amount of time required to train the model and make predictions. A shorter execution time indicates that the model is more efficient, which can be particularly important in real-world applications where time constraints are critical. By evaluating the execution time, we can assess the trade-offs between model performance and computational resources. These evaluation metrics provide a comprehensive and robust assessment of our neural network and proposed model's performance. By considering multiple aspects of performance, we can ensure that our model is not only accurate but also efficient, generalizable, and reliable across various datasets and application scenarios.

## V. RESULT AND DISCUSSION

In this study, we investigated the role of semantic and syntactic features in vulnerability prediction for CWE-119, focusing on buffer overflow vulnerabilities. We began by converting the text dataset into a minimal intermediate representation using a tokenizer provided by the Keras library. This basic representation assigns a numerical value to each word without considering semantic information. Since the meaning of code is often better captured by considering the context of multiple words, we employed state-of-the-art word embedding algorithms—GloVe and fastText—to extract semantic features from function-level codes. These features were then fed into neural network models for vulnerability prediction. We used 100,000 instances for training, 20,000 for validation, and 10,000 for testing. Our evaluation metrics included accuracy, precision, recall, and F1 score, with a focus on minimizing false positives and false negatives. We trained seven neural network models (Simple RNN, LSTM, BiLSTM, word2vec, BERT, GPT-2, and LSTM-Autoencoder) and our proposed stacking ensemble neural network model. Our ensemble learning model outperformed single models, achieving the highest accuracy in vulnerability prediction.

Table 2 presents the results of vulnerable source code classification using different neural network models without word embedding algorithms. The Simple RNN model achieves an accuracy of 0.89, precision of 0.88, recall of 0.88, and F1 score of 0.92, with an execution time of 42 minutes and 8 seconds. The LSTM model has slightly better performance with an accuracy of 0.90, precision of 0.90, recall of 0.90, and F1 score of 0.92, and takes 29 minutes and 48 seconds to run. The BiLSTM model shows further improvement, obtaining an accuracy of 0.91, precision of 0.93, recall of 0.90, and F1 score of 0.87, but requires 2 hours and 5 minutes for execution.

The Word2vec model yields an accuracy of 0.89, precision of 0.92, recall of 0.95, and F1 score of 0.93, with a runtime of 40 minutes and 2 seconds. The LSTM Autoencoder model has an accuracy of 0.91, precision of 0.93, recall of 0.94, and F1 score of 0.94, taking 53 minutes and 13 seconds for execution. The BERT model performs better with an accuracy of 0.92, precision of 0.93, recall of 0.93, and F1 score of 0.95, but requires 2 hours and 38 minutes to run. The GPT-2 model has an accuracy of 0.92, precision of 0.97, recall of 0.98, and F1 score of 0.97, with a considerably longer execution time of 7 hours and 48 minutes. Lastly, the proposed model outperforms the other models with an accuracy of 0.94, precision of 0.99, recall of 0.98, and F1 score of 0.99, and takes 2 hours and 31 minutes to execute.

Table 3 shows the results when using GloVe and FastText embeddings. In general, the performance of all models improved when using these embeddings. The Simple RNN, LSTM, BiLSTM, and Word2vec models show a similar trend in improvement, with their respective accuracies increasing to 0.92, 0.92, 0.93, and 0.94. The LSTM Autoencoder model's performance slightly decreased with an accuracy of 0.90. The BERT, GPT-2, and proposed models maintain their superior performance, with accuracies of 0.94, 0.95, and 0.95, respectively. The execution times for all models vary, with the proposed model having a runtime of 2 hours and 46 minutes.

Figure 3 shows the performance metrics for different neural network models on vulnerable source code without using any word embedding algorithms. The models considered are Simple RNN, LSTM, BiLSTM, Word2vec, LSTMAutoencoder, BERT, GPT-2, and a proposed model. The metrics considered are accuracy, precision, recall, and F1 score. The results demonstrate that the proposed model outperforms all other models in terms of accuracy and F1 score, achieving an accuracy of 0.94 and an F1 score of 0.99. The execution time of the proposed model is also relatively fast compared to other models, taking only 2 hours and 31 minutes.

Figure 4 presents the classification results of the same neural network models on vulnerable source code using GloVe and fastText word embedding algorithms. The results demonstrate that all models achieved higher accuracy and F1 score compared to the results in Figure 3. The proposed model continues to perform the best with an accuracy of 0.95 and an F1 score of 0.99. However, the execution time of the proposed model is longer compared to Figure 3, taking 2 hours and 46 minutes.

These figures provide a clear comparison of the performance of different neural network models and highlight the effectiveness of using word embedding algorithms for improving the classification results of vulnerable source code. The proposed model performs well in both scenarios, showing its potential as a reliable classification model.

In Table 4, we present a comparison analysis between our proposed model and previous works in the domain of vulnerability detection. The table highlights the differences in terms of the purpose of each study, the data used, whether semantic or syntactic feature extraction was performed, the highest performance achieved, and if efficiency measurements were conducted.

Lorga et al. [18] aimed at vulnerability detection using Twitter text data, but they did not perform semantic or syntactic feature extraction. Their model achieved an accuracy of 94.96%, and they did not provide any efficiency measurements. Similarly, Foret et al. [39] worked on vulnerability detection using news articles without incorporating semantic or syntactic features, resulting in an 87% accuracy. No efficiency measurement analysis was conducted in their work as well. Harer et al. [20] and Russell et al. [22] both focused on vulnerability detection in source code but did not consider semantic or syntactic feature extraction. Their models achieved F1-scores of 49.99% and 56.6%, respectively, without any efficiency measurement analysis. Behzdan et al. [40] also worked on vulnerability detection in source code without extracting semantic or syntactic features. They reported an accuracy of 94.72%, but no efficiency measurement analysis was performed.

Our proposed model targets vulnerability detection in source code and incorporates semantic and syntactic feature extraction using GloVe and fastText embeddings. As a result, our model achieves the highest accuracy of 95% compared to the previous works. Moreover, we contribute to efficient measurement analysis and perform an in-depth analysis of the features that were not considered in previous studies. This comprehensive approach allows us to better understand the factors influencing the performance of vulnerability detection models and develop more effective methods for detecting security vulnerabilities in source code.

TABLE II: Vulnerable Source code Classification results using different Neural network models with no word embedding algorithms

| Models | Accuracy | precision | Recall | F1 Score | Execution Time |
|---|---|---|---|---|---|
| Simple RNN | 0.89 | 0.88 | 0.88 | 0.92 | 42min 8s |
| LSTM | 0.90 | 0.90 | 0.90 | 0.92 | 29min 48s |
| BiLSTM | 0.91 | 0.93 | 0.90 | 0.87 | 2h 5min |
| Word2vec | 0.89 | 0.92 | 0.95 | 0.93 | 40min 2s |
| LSTMAutoencoder | 0.91 | 0.93 | 0.94 | 0.94 | 53min 13s |
| BERT | 0.92 | 0.93 | 0.93 | 0.95 | 2h 38min |
| Gpt2 | 0.92 | 0.97 | 0.98 | 0.97 | 7h 48min |
| Proposed Model | 0.94 | 0.99 | 0.98 | 0.99 | 2h 31min |

TABLE III: Vulnerable Source code Classification results using different Neural network models with embedding algorithms GloVe + fastText

| Models | Accuracy | precision | Recall | F1 Score | Execution time |
|---|---|---|---|---|---|
| Simple RNN | 0.92 | 0.93 | 0.93 | 0.97 | 42min 8s |
| LSTM | 0.92 | 0.93 | 0.95 | 0.97 | 33min 13s |
| BiLSTM | 0.93 | 0.96 | 0.96 | 0.99 | 45min 3s |
| Word2vec | 0.94 | 1.00 | 0.98 | 0.99 | 42min 56s |
| LSTMAutoencoder | 0.90 | 0.93 | 0.94 | 0.95 | 59min 53s |
| BERT | 0.94 | 0.95 | 0.95 | 0.99 | 5h 16min |
| Gpt2 | 0.95 | 0.97 | 0.98 | 0.99 | 8h 33min |
| Proposed Model | 0.95 | 0.97 | 0.98 | 0.99 | 2h 46min |

TABLE IV: Comparative analysis with previous work

| Previous authors | Purpose | Data | Semantic or Syntactic feature extraction? | Highest percentage | Efficiency Measurement? |
|---|---|---|---|---|---|
| Lorga et al. [18] | Vulnerability detection | Twitter text data | No | 94.96% (Accuracy) | No |
| Foret et al. [39] | Vulnerability detection | News Articles | No | 87% (Accuracy) | No |
| Harer et al. [20] | Vulnerability detection | Source code | No | 49.99% (F1-score) | No |
| Russell et al. [22] | Vulnerability detection | Source code | No | 56.6% (F1-score) | No |
| Behzadan et al. [40] | Vulnerability detection | Source code | No | 94.72% (Accuracy) | No |
| **Our Proposed Model** | **Vulnerability detection** | **Source code** | **Yes** | **95%** (Accuracy) | **Yes** |

## VI. CONCLUSION

Our research aims to detect implementation vulnerabilities early in the development cycle by leveraging the power of neural networks. We have collected a large dataset of open-source C and C++ code and developed a scalable and efficient vulnerability detection method based on various neural network models. We compared the performance of different models, including Simple RNN, LSTM, BiLSTM, LSTM-Autoencoder, Word2Vec, BERT, and GPT-2, and found that models with semantic and syntactic information extracted using state-of-the-art word embedding algorithms such as GloVe and FastText outperform those with a minimal text representation. Our proposed neural network model has shown to provide higher accuracy with greater efficiency than the other models evaluated. We have also analyzed the execution time of the models and proposed a trade-off between accuracy and efficiency. Overall, our research contributes to the development of large-scale machine learning systems for function-level vulnerability identification in source code auditing.

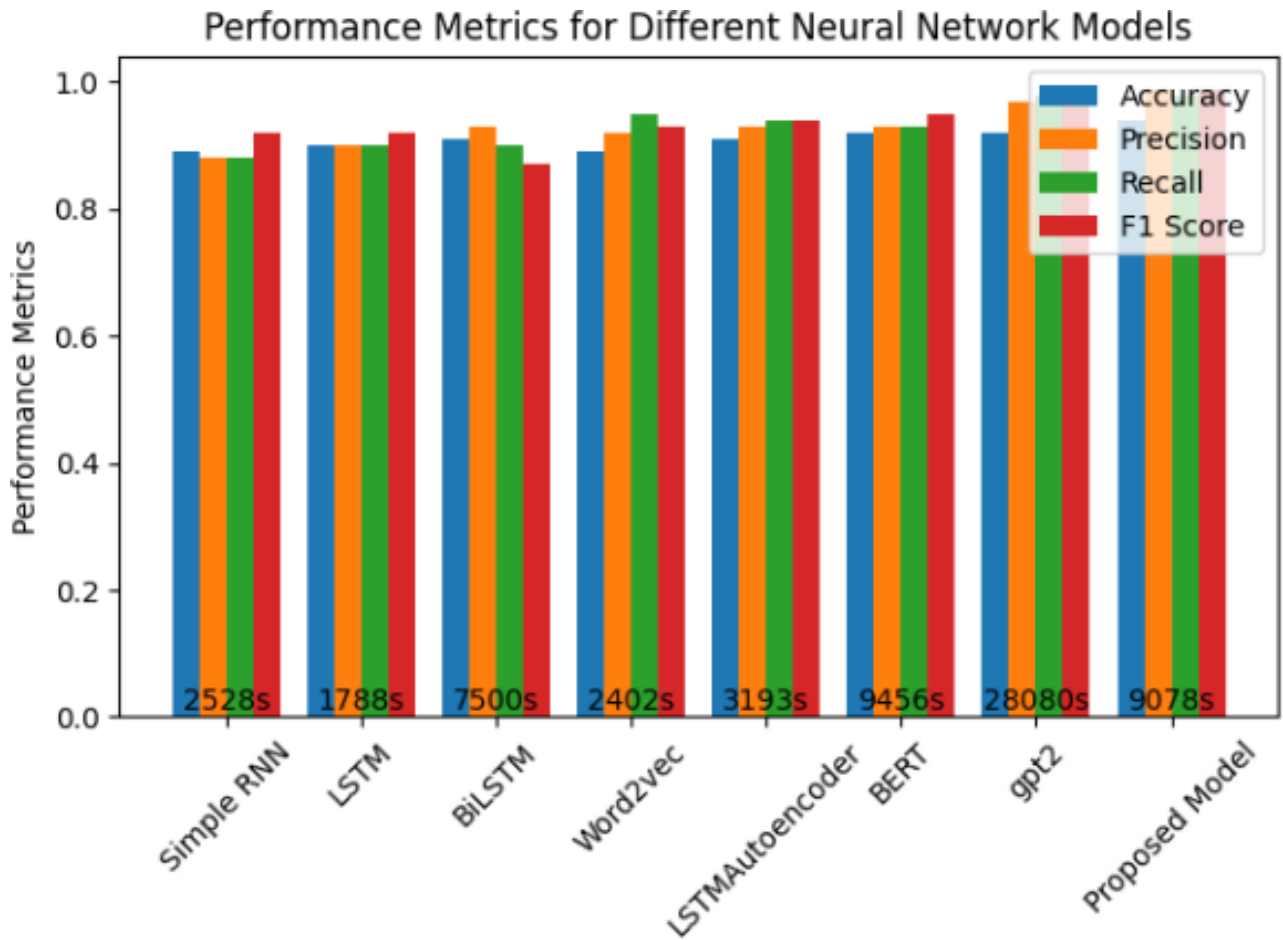

Fig. 3: Performance Metrics for Different Neural Network Models on Vulnerable Source Code without Word Embedding Algorithms

ACKNOWLEDGEMENT

The work is supported by the National Science Foundation under NSF Award #2209638, #2100115, #2209637, #2100134, #1663350. Any opinions, findings, recommendations, expressed in this material are those of the authors and do not necessarily reflect the views of the National Science Foundation.

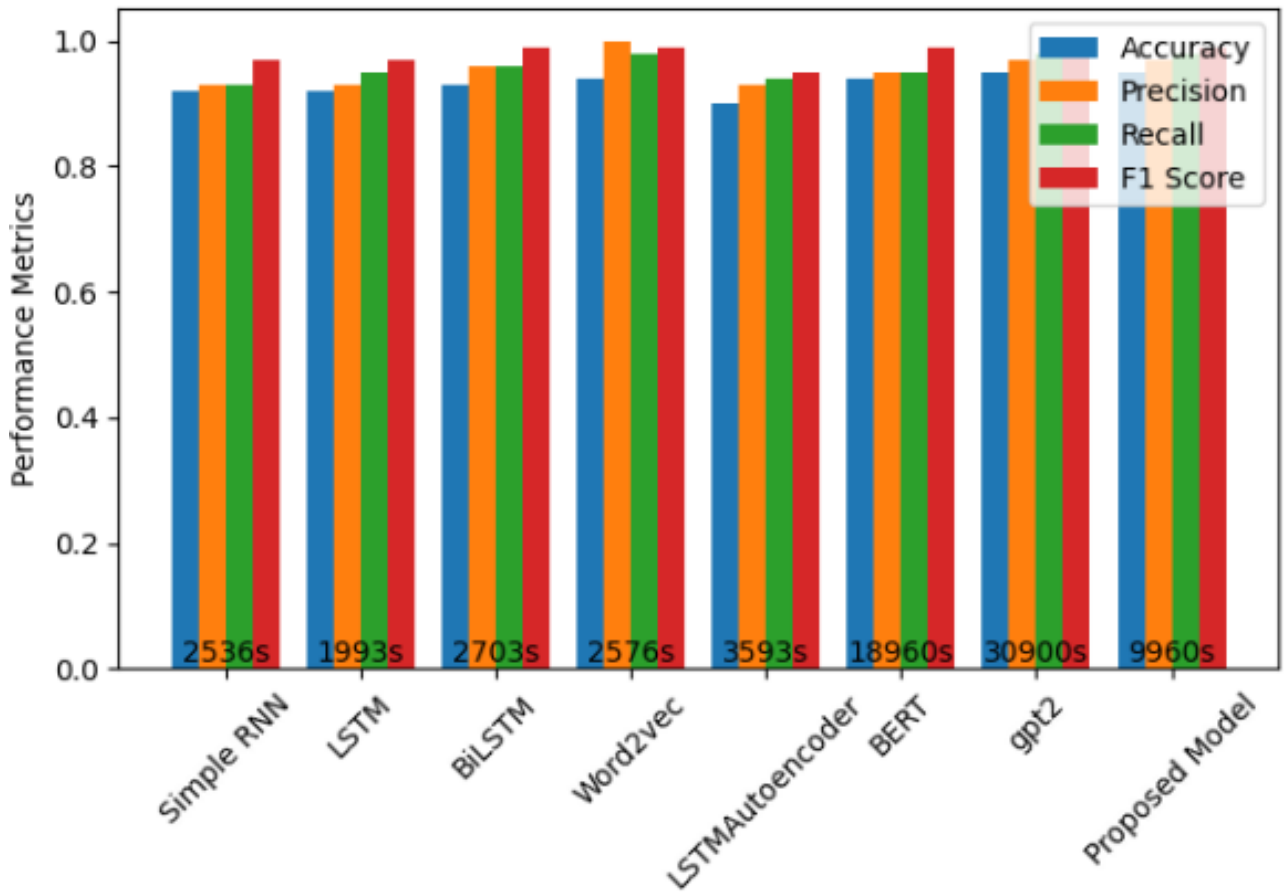

Fig. 4: Performance Metrics for Different Neural Network Models on Vulnerable Source Code without Word Embedding Algorithms